%% file: main.tex
\pdfoutput=1
\documentclass{article}
\usepackage[preprint]{spconf}
\usepackage{spconf,amsmath,graphicx,hyperref}

\usepackage{algorithmic}
\usepackage{graphicx}
\usepackage{textcomp}
\usepackage{xcolor}
\usepackage{siunitx}
\usepackage{amsfonts}
\usepackage{makecell}
\usepackage[symbol]{footmisc}
\usepackage{tabularx}
\usepackage{pifont}
\usepackage{booktabs}
\usepackage[capitalize]{cleveref}
\usepackage{tikz}
\usepackage{pgfplots}
\pgfplotsset{compat=1.18}
\usepackage{bm}
\usepackage{balance}
\usepackage{upgreek}
\usepackage{multirow} 
\usetikzlibrary{positioning, fit, calc}
\usepackage[acronym, shortcuts, toc,nonumberlist]{glossaries}
\usepackage{cite}
\usepackage{xcolor}
\usepackage{twemojis}
\usepackage{amsmath}

\usetikzlibrary{
    positioning,
    calc,
    arrows.meta,
    shadows
}

\tikzset{
    >=stealth,
    arrow/.style={->, line width=0.1em},
    line/.style={-, line width=0.1em},
    reverse arrow/.style={<-,shorten <=0.1em}
}

\def\tsix{{$\text{T}_{60} $ }}
\def\tsixnospace{{$\text{T}_{60}$}}

\title{ROOM IMPULSE RESPONSE EMBEDDINGS FOR SPEECH ENHANCEMENT \\ IN NOISY AND REVERBERANT ENVIRONMENTS}

\name{Adrian Meise, Reinhold Haeb-Umbach}

\address{Paderborn University, Germany}

\input{macros}

\copyrightnotice{\parbox{\textwidth}{This work has been submitted to the IEEE for possible publication.\\ Copyright may be transferred without notice, after which this version may no longer be accessible.}}

\begin{document}
\ninept

\setlength{\textfloatsep}{8pt plus 0.0pt minus 0.0pt}
\setlength{\dbltextfloatsep}{11pt plus 2pt minus 4pt}

\maketitle

\begin{abstract}
We propose a self-supervised approach for learning \gls{RIR} representations from single-channel noisy-reverberant speech. It consists of first training on reverberant data, then on noisy-reverberant data, and finally with a teacher-student approach, where the student learns to replicate the teacher's embeddings when given a noisy version of the reverberant input. We assess their representational capabilities by estimating acoustic room parameters from them. Conditioning a discriminative speech enhancement model on the derived embeddings yields consistent gains across all evaluated metrics, including downstream word error rate, for both reverberant and noisy-reverberant speech.
\end{abstract}
\section{Introduction}
\label{sec:intro}
\glsresetall
Speech dereverberation is beneficial both for human perception and for downstream tasks such as \gls{ASR}.
Both for earlier model-based approaches and today's data-driven methods, it has been shown that using or estimating side information about the spatial signal propagation characteristics can help to improve dereverberation performance \cite{kodrasi_14_filtering, wu_2017_t60, wang21l_interspeech, li_2023_t60, bahrman_26_sampling}. %
Such information can be in the form of a \gls{RIR} or its short-time frequency-domain counterpart, the convolutive transfer function, or in the form of sound propagation parameters, such as sound decay times like \tsixnospace. This enables the model to better account for the physics of the reverberant acoustic environment.

An alternative to estimating signal propagation characteristics with supervised learning algorithms or even assuming knowledge of them is the use of \gls{SSL} techniques. They have been very successful for capturing spectral signal properties, while spatial SSL representations are less well-researched. 
In \cite{goetz_23_ssl_parameters}, representations from simulated reverberant input signals are extracted via contrastive learning and used for an acoustic parameter estimation task.
Similarly, \cite{yang_24_generative} chose a predictive pretext task for \gls{SSL} to capture universal spatial acoustic representations from two-channel input with a high \gls{SNR}, which are evaluated on an acoustic parameter estimation task as well.
But also for room classification \cite{bitterman24_interspeech}, \gls{RIR}
estimation \cite{av_rir_24} and for the generation of new \glspl{RIR} \cite{lluis_25_rir_gen}, embeddings extracted from an encoder trained in a contrastive framework have been proposed.

The usefulness of spatial SSL representations for speech enhancement tasks is less investigated. 
In \cite{khokhlov19_interspeech}, RIR embeddings are used to improve ASR performance for degraded input signals, where self-supervision is achieved by predicting the indices of the RIR that generated reverberant input.
Khanagha et al. showed that discriminative U-Net-based dereverberation models learn implicit representations of the \gls{RIR}  \cite{khanagha_2026}. Based on this observation, it was proposed to condition a diffusion-based model on pretrained embeddings that are learnt in a contrastive fashion. Conditioning the model on these \gls{RIR} embeddings allowed for reducing the number of diffusion steps while achieving improved speech enhancement scores compared to using no conditioning.  However, the analysis only considered a generative, diffusion-based dereverberation approach.

In a realistic recording environment with a distant microphone, the recorded signal will be affected not only by multi-path propagation, but also by additive acoustic distortions, such as noise. In general, robustness of \gls{SSL} features to noise or other perturbations is an ongoing research question and open challenge, both for speech \cite{song_2024} and audio processing \cite{fujimura_2026}. For \gls{RIR} representations, this is also shown in \cite{xiang2026_rir}, where the uncertainty of \gls{RIR} embeddings derived with a variational autoencoder is investigated, showing that noise leads to representation shift and embedding dispersion.

In this work, we investigate the learning of \gls{RIR} representations in the presence of noise in challenging \gls{SNR} conditions. We propose to directly infer self-supervised representations from noisy and reverberant input using a curriculum learning approach. In the first stage, following \cite{khanagha_2026}, RIR embeddings are learnt from single-channel reverberant speech with a contrastive loss. In the second training stage, contrastive learning is continued, however, with noisy-reverberant speech input,  while in the third stage a teacher-student approach is used, where the student reproduces the teacher's embeddings from reverberant input using the corresponding noisy-reverberant input.
Evaluation on acoustic parameter estimation tasks demonstrates the representation capabilities of the noise-robust embeddings. Further, we assess the impact on speech enhancement of noisy-reverberant data, showing that the conditioning of a discriminative deep filtering-based enhancement model on those embeddings leads to consistent improvements for both reverberant and noisy-reverberant data.
   
In the next section, we explain the training stages.  \Cref{sec:setup} summarizes experimental details, and \Cref{sec:results} evaluates the representation quality of the embeddings, before the application for speech enhancement is evaluated in \Cref{sec:speech_enhancement}.

\section{SSL RIR representation learning}
\label{sec:main}
Speech enhancement in practical applications is not only confronted with reverberation, but also with additive noise signals emerging from the environment. 
Here, we aim to learn embeddings from noisy and reverberated input that reflect the information concerning the \gls{RIR}, while the noise and source signal should be ignored.
Therefore, noise-robust representations of the \gls{RIR} should be derived that exhibit the same characteristics as embeddings obtained from reverberant-only input. 
To achieve this, one option is to first perform denoising of the input and then to derive the embedding from the denoised signal. However, this assumes that the reverberation is transparent to the denoising system, which is typically not the case \cite{itg_paper}. Consequently, possible artifacts generated by the enhancement system will impact the training of the \gls{RIR} encoder, and the computed embeddings will become dependent on the denoising system.

To facilitate native noise-robust SSL, we adopt a three-stage training approach, which will be detailed below and which can be summarized as follows:
 \textit{1. Training on reverberant data}, \textit{2. Training on noisy-reverberant data} and \textit{3. Teacher-student training} with the teacher from stage 1 and the student from stage 2.

\subsection{Contrastive learning of RIR representations}
\label{sec:rir_encoder}
The first stage is adopted from \cite{khanagha_2026}. There, it is proposed to improve diffusion-based dereverberation via pretrained \gls{RIR}-dependent representations obtained from an \gls{RIR} encoder that is trained in a self-supervised fashion. 
The self-supervision is achieved via a contrastive learning task, where positive and negative pairs of the input data are generated.
A positive pair consists of two different speech utterances that are convolved with the same room impulse response. These are encouraged to produce similar embeddings that are close in the embedding space, as they correspond to the same \gls{RIR}, while the source signal is to be ignored. Consequently, samples belonging to different \glspl{RIR} should be mapped to distinct regions.
In the negative pair, the source utterances are the same, such that the input mixtures differ only in their spatial properties in terms of different \glspl{RIR}, which should be pushed apart. Therefore, these negative pairs should explicitly encourage the model to ignore the source utterance and focus only on the reverberation characteristics caused by the \gls{RIR}.
Considering two utterances $u_1$ and $u_2$ and two room impulse responses $h_1$ and $h_2$, three relevant mixture types can be distinguished: $s_1 = u_1 \ast h_1$, $s_2 = u_2 \ast h_1$ and $s_3 = u_1 \ast h_2$, where $s_1$ and $s_2$ would be a positive pair and $s_1$ and $s_3$ would be a negative pair.
The loss function consists of a weighted sum of the InfoNCE loss \cite{infonce_2019} and cosine similarity for negative pairs.

\subsection{Noise-robust RIR representation learning}
\label{sec: stage2}
To derive embeddings from noisy-reverberant input, we propose to continue training the model from the previous stage on noisy-reverberant data with the same contrastive learning framework as before. This way, the system adapts to noise while building upon the pretrained capabilities already learned for less challenging reverberant input, where the desired structure of the embedding space is easier to obtain than when training from scratch on noisy-reverberant input.
We encourage selecting different noise samples for the same \gls{RIR} in the positive pair, so that the same \gls{RIR} has to be identified despite different noise samples.
For the negative pair with different \glspl{RIR}, the noise is instead sampled to be the same with a higher probability. 
As with the source signal, the model is encouraged to ignore the noise and to learn that the difference between the samples lies in the different \glspl{RIR}. This prevents the model from identifying different RIRs based on different noise samples.
For the three considered mixtures $s_1$, $ s_2, s_3$ and three different noise samples $n_1, n_2, n_3$, this can be summarized as 
    \begin{align*}
        s_1 &= u_1 \ast h_1 + n_1\\[-10pt]
        s_2 &= u_2 \ast h_1  + n \quad \text{with} \quad  n = \begin{cases}
            n_2 \text{ with prob. } p \\[-1pt]
            n_1 \text{ with prob. } 1-p
        \end{cases} \\[-2pt]
        s_3 &= u_1 \ast h_2 + n  \quad \text{with} \quad  n = \begin{cases}
            n_1 \text{ with prob. } p\\[-1pt]
            n_3 \text{ with prob. } 1-p,
        \end{cases} 
    \end{align*}
where $p=0.7$ is chosen in the following.

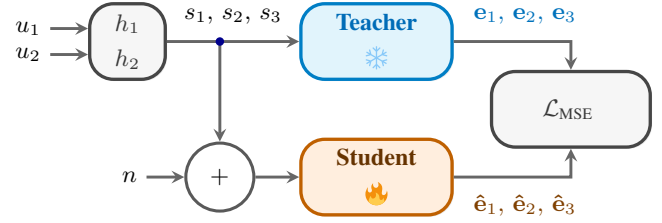
\begin{figure}
    \centering
    \input{images/teacher_student}
    \caption{Overview of the proposed teacher-student training, where the frozen teacher from stage 1 extracts embeddings from the reverberant input and the student from stage 2 has to replicate the embedding from the noisy and reverberant input.}
    \label{fig:teacher_student}
\end{figure}

\subsection{Teacher-student training}
Following the previous stage, we propose a third stage of teacher-student training to adapt the embedding space learnt on noisy-reverberant data to the embedding space learnt from only reverberant data, see \Cref{fig:teacher_student}.
Here, the embeddings extracted from the reverberant-only input by the frozen teacher from stage 1 serve as the target, which should be replicated by the student from stage 2, which receives noisy and reverberant input. The teacher and student use the same model architecture. Therefore, the characteristics of the teacher embedding space are imprinted onto the student to avoid the pollution of the embedding space with noise information. The setup of positive and negative pairs among $s_1, s_2$ and $s_3$ is maintained as in previous stages. 
Here, the reverberant mixture is fed into the frozen teacher model, which predicts the embedding $\mathbf{e}_i$ from input $s_i$, while the noisy-reverberant mixture is input to the student model, which outputs the embedding $\mathbf{\hat{e}}_i$, and the cost function in terms of the mean squared error (MSE) $\mathcal{L}_\text{MSE} = \|\mathbf{e}_i -\mathbf{\hat{e}}_i\|^2$ is used to train the student. 
For validation, we keep the contrastive loss between the derived embeddings as the selection criterion, since it measures the desired characteristics of the embedding space, while the MSE loss is only a means to achieve this.

\section{Experimental setup}
\label{sec:setup}

For the experiments, we use data from the EARS \cite{richter24_interspeech} dataset with 100 hours of clean speech and the corresponding variants EARS-WHAM v2 and EARS-Reverb v2. 
For the \gls{RIR} encoder, we select the Conformer-based \cite{conformer} architecture, as it achieved the best results among the proposed architectures  
in \cite{khanagha_2026}. 
It uses the log-magnitude of the reverberant signal as input and consists of 10 layers with 256 hidden units and 4 attention heads, and a two-layer MLP head that produces 256-dimensional $\ell_2$-normalized embeddings.
We start training from the pretrained checkpoint\footnote{https://github.com/
sp-uhh/rir-encoder}
to leverage the self-supervised pretraining, and we fine-tune the model on the EARS-Reverb data in the first stage to adapt to the EARS source signal data, which exhibit a broader dynamic range, including emotional and highly dynamic speech. 
In the second stage, we add noise from the WHAM! \cite{wichern19_interspeech} database, which has a total duration of 80 hours. Here, noise is added with an SNR of \SIrange{-2.5}{17.5}{\dB} to the EARS-Reverb data, effectively generating noisy-reverberant EARS-WHAMR data \cite{itg_paper}. In the third teacher-student training stage, we additionally use noise from CHiME-3 \cite{chime3}, SINS \cite{sins}, and SMS-WSJ \cite{smswsj_2019} for a wider variety of noise samples and more robust noise performance.

To evaluate generalization to unseen noise types and rooms, we generate new degraded test sets from the clean EARS test split using noise and RIRs not seen during training of the models.
For the \textit{reverberant} data, we select measured \glspl{RIR} from the BUT-ReverbDB \cite{but_rir} database. 
For the \textit{noisy-reverberant} version, we add  noise from the DEMAND \cite{demand_2013} database with an \gls{SNR} from \SIrange{0}{20}{\dB}.

In order to gain insight into what has been learnt by the \gls{RIR} embeddings, we train simple estimators that jointly estimate \tsix and \gls{DRR}. Since these parameters are sufficient to sample a \gls{RIR} \cite{bahrman_26_sampling}, they are likely to be represented in the estimated embeddings in order to discern different \glspl{RIR}. On the one hand, this allows the evaluation of the representation capabilities and an assessment of whether room-related parameters are encoded. 
On the other hand, it is investigated whether the learnt student embeddings represent the same information as the teacher embeddings. 
We use a simple linear layer with 514 parameters to estimate \tsix and DRR from the derived embeddings, where an individual estimator is trained for the embeddings at each training stage.
This estimator is deliberately chosen to be lightweight, so that estimation performance is dependent on the representation capabilities instead of a powerful estimator. For comparison, a 1D-CNN-based estimator with 270,210 parameters is trained that directly operates on the waveform input. Results are reported in terms of \gls{MAE} and Spearman correlation coefficient $\rho$ to measure the monotonicity of the relationship for both estimated parameters.

\section{Evaluation}
\label{sec:results}

\begin{figure*}[t]
\centering
\input{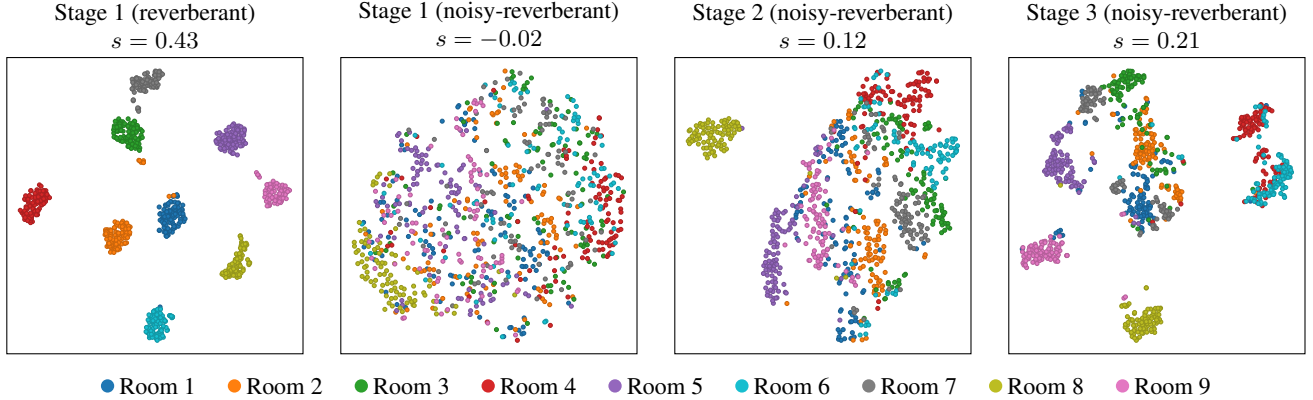}
\caption{Comparison of t-SNE projections and Silhouette Coefficients $s$ of  \gls{RIR} embeddings for reverberant and noisy-reverberant data after stage 1, after training on noisy and reverberant data in stage 2, and after training as the student network (from left to right).}
\label{fig:tsne-fig}
\end{figure*}

\begin{table}[t]
    \centering
    \caption{Performance of acoustic-parameter estimation from the raw
    waveform or from the embeddings after each training stage. The number of parameters is denoted in parentheses. Bold results indicate best performance for each metric.  The MAEs for \tsix and DRR are reported in seconds and \si{\dB}, respectively.}
    \vspace{3pt}
    \setlength{\tabcolsep}{4.5pt}
    \begin{tabular}{l l c c c c }
        \toprule
        Test condition & Model & Embedding & Metric & MAE $\downarrow$ & $\rho$ $\uparrow$\\
        \midrule

        \multirow{9}{*}{\makecell{Reverb-\\erant}}
        & \multirow{2}{*}{\makecell{CNN \\ (0.27M)}}
        & \multirow{2}{*}{--}
        & \tsix & 0.30 & 0.45 \\
        & & & DRR & 7.36 & 0.28 \\
        \cmidrule(lr){2-6}

        & \multirow{7}{*}{\makecell{Linear \\ (514)}}
        & \multirow{2}{*}{\makecell{Stage 1 \\ (Teacher)}}
        & \tsix &  0.18 & \bfseries 0.76 \\        
        & & & DRR & \bfseries 2.51 &   0.52 \\
        \cmidrule(lr){3-6}

        & & \multirow{2}{*}{Stage 2}
        & \tsix & \bfseries 0.17 & 0.74 \\
        & & & DRR &  2.74 & \bfseries 0.56 \\
        \cmidrule(lr){3-6}
        
        & & \multirow{2}{*}{\makecell{Stage 3 \\ (Student)}}
        & \tsix & 0.21 & 0.73 \\
        & & & DRR &  2.65 & \bfseries  0.56 \\

        \midrule

        \multirow{12}{*}{\makecell{Noisy-\\reverberant}}
        & \multirow{2}{*}{\makecell{CNN \\(0.27M)}}
        & \multirow{2}{*}{--}
        & \tsix & 0.31 & 0.28 \\
        & & & DRR & 5.26 & 0.21 \\
        \cmidrule(lr){2-6}

        & \multirow{10}{*}{\makecell{Linear \\ (514)}}       
        & \multirow{2}{*}{\makecell{Stage 1 \\ (Teacher)}}
        & \tsix & 0.25 & 0.46 \\
        & & & DRR & 3.37 & 0.24 \\
        \cmidrule(lr){3-6}

        & & \multirow{2}{*}{\makecell{Denoised\\ + Stage 1}}
        & \tsix & 0.27 & 0.32 \\
        & & & DRR & 4.02 & 0.21 \\
        \cmidrule(lr){3-6}

        & & \multirow{2}{*}{Stage 2}
        & \tsix & \bfseries 0.22 &  0.67 \\
        & & & DRR & 3.66 & 0.38 \\
        \cmidrule(lr){3-6}
        
        & & \multirow{2}{*}{\makecell{Stage 3 \\ (Student)}}
        & \tsix &  0.23 & \bfseries 0.68 \\
        & & & DRR &  \bfseries 2.56 & \bfseries 0.51 \\

        \bottomrule
    \end{tabular}
    \label{tab:parameters}

\end{table}

The results for \tsix and DRR estimation performance are shown in \Cref{tab:parameters}. The evaluation on reverberant input indicates that the derived embeddings contain relevant information to estimate \tsix and DRR. All embedding-based models achieve better parameter estimates compared to the CNN estimator, indicating that relevant features for the parameter estimation are more readily accessible from the embeddings.
When comparing the representations from the embedding models at different training stages, it can be observed that although the models from stages 2 and 3 are trained on noisy-reverberant input, they still contain information for reverberant-only test input.
Note that the performance of this parameter estimation task could be improved by using more advanced estimators. However, for incorporation of the embeddings into downstream models,
FiLM \cite{film} layers are often used for conditioning, where the embeddings are also processed by small networks.
    
On noisy-reverberant data, the model from stage 1 fails to represent \tsix and DRR information and achieves performance not much better than the CNN-based estimator. For comparison, we also first denoised the input with \cite{schroter23b_interspeech} before extracting the embeddings from the enhanced signals, but reverberation is not transparent to the denoising system, and correspondingly, the embeddings no longer contain relevant features.
In contrast, the embeddings estimated by the model after the second stage contain relevant information on \tsix and \gls{DRR}. The student model from stage 3 is able to further increase estimation performance, especially on DRR estimation, outperforming the stage 2 model by more than \SI{1}{\dB} and achieving a correlation coefficient similar to the results obtained from reverberant input.

In \Cref{fig:tsne-fig}, we provide a qualitative comparison of the learnt embedding spaces at the different training stages. 
For clearer visualization, we use a modified version of the test dataset with only one \gls{RIR} sampled per room, given the wide variety of recording setups for each room in \cite{but_rir}.
The resulting embeddings are projected onto two dimensions using t-SNE \cite{tsne}.
Additionally, Silhouette Coefficients \cite{silhouettes_87} $s \in [-1, 1]$ are reported, where 1 indicates the best result, which is reduced for overlapping clusters or wrong assignments.
For the reverberant data, the embeddings from stage 1 form nine distinguishable clusters corresponding to the nine rooms occurring in the test data. However, if this model is directly applied to noisy-reverberant data, no regularized embedding space is generated, and the grouping of embeddings of the same \gls{RIR} into similar regions in the embedding space is lost, highlighting the need for noise-robust representations. After stage 2 of the training, embeddings belonging to the same room are placed closer together in the embedding space, but only the samples from room 8 form a cluster as desired, while the embeddings of one \gls{RIR} scatter and intersperse with embeddings belonging to other \glspl{RIR}.
For the embeddings derived by the student network, distinguishable clusters that are also separate from clusters belonging to other rooms can again be identified. Here, the most structured embedding space is achieved, which resembles the desired embedding space of the teacher for reverberant input the most. 

\section{Application to speech enhancement}
\label{sec:speech_enhancement}

\subsection{Deep filtering}
In \cite{schroter23b_interspeech}, the lightweight and real-time-capable speech enhancement system DeepFilterNet (``DFN3'') was proposed. Here, a filter vector is predicted and applied to the degraded input to obtain an enhanced output. Since it was originally proposed for denoising, its dereverberation capabilities are limited \cite{rosenbaum_2025}, and we thus expect performance gains from additional conditioning on features that represent \gls{RIR} information.
In \cite{rosenbaum_2025}, it is proposed to extend the functionality by introducing an additional second dereverberation step that estimates the late reverberation by filtering the delayed input vector  (``two-step''). Here, we suspect that  additional \gls{RIR} features are promising for the estimation of these filter coefficients, to provide a better estimation of the late reverberation and better dereverberation.

For the training, the parameters and training setup are chosen as described in \cite{rosenbaum_2025}. To condition these models on the \gls{RIR} embeddings, we incorporate FiLM \cite{film} layers after every block in the decoder parts of the models. The embeddings are input to the FiLM layers, which estimate feature-wise scaling and shifting parameters with a small MLP.
Thus, the embeddings can impact the estimation of the filter coefficients. 

We evaluate performance on the generated reverberant and noisy-reverberant datasets.
For the embeddings derived from reverberant data, the \gls{RIR} encoder after the first training stage is used, and for noisy-reverberant data, the embeddings are obtained from the student after training stage 3.
We select a set of metrics that encompasses two intrusive measures,  STOI \cite{stoi} and cepstral distance (CD) \cite{cd}, as well as two non-intrusive ones, speech-to-reverberation modulation energy ratio (SRMR) \cite{srmr}, and DNSMOS \cite{dnsmos_21}. Additionally, we report the \gls{WER} as an indicator of speech (machine) intelligibility. 
For \gls{ASR}, we employ  a lightweight QuartzNet-based model (18.9M param.) and a large Conformer-based model (1.1B param.) from the NeMo toolkit \cite{nemo}. We use two models to check if performance improvements are obtained not only for a weak but also for a strong, potentially more robust recognizer.

\begin{table}[ht]
\vspace{-2.1mm}
    \centering
    \caption{Performance for reverberant and noisy-reverberant datasets for models with and without additional conditioning. Bold results indicate the best performance per model type. WER results are reported for the weak (left/) and strong (/right) ASR model.}
\vspace{3pt}
\sisetup{table-format=1.2+-1.2}
\sisetup{
text-series-to-math = true,
propagate-math-font = true
}
\setlength{\tabcolsep}{2pt}

    \begin{tabular}{l H c H c c c c }
    \toprule
         \multirow{2}{*}{Model} & \multicolumn{7}{c}{Reverberant}\\
         \cmidrule(lr){2-8}
         & {PESQ} & {STOI $\uparrow$} & {FSNR} & {CD $\downarrow$} &  {SRMR $\uparrow$} & {DNSMOS $\uparrow$} & {WER $\downarrow$} \\
         \midrule
         Input & 1.24 &	0.64 &	0.97 &	4.30 &	3.77 &	2.87 &	36.85 / 12.05	\\  
         \midrule
         
         DFN3 \cite{schroter23b_interspeech} & 1.28 &	0.72 &	\bfseries 2.80 &	3.75 &	6.18 &	2.80 &	44.19  / 12.88  \\ 
        
          \multicolumn{1}{r}{+ FiLM} & \bfseries 1.31 &	\bfseries 0.73 &	2.19 &	\bfseries 3.64 &	\bfseries 6.20 &	\bfseries 2.84 &	\bfseries 39.52	/ 10.56  \\  
          
         \midrule
         Two-step \cite{rosenbaum_2025} & 
          \bfseries 1.38 &	0.73 &	2.29 &	3.81 &	7.41 &	2.83 &	42.91 / 12.06  \\
         \multicolumn{1}{r}{+ FiLM}
         & 1.33 &	\bfseries 0.74 &	\bfseries 2.56 &	\bfseries 3.55 &	\bfseries 8.39 &	\bfseries 2.85 &	\bfseries 38.81	/ 11.13  \\  

        \bottomrule
    \end{tabular}

    \par\bigskip

    \begin{tabular}{l H c H c c c c }
    \toprule
         \multirow{2}{*}{Model} & \multicolumn{7}{c}{Noisy-reverberant}\\
         \cmidrule(lr){2-8}
         & {PESQ} & {STOI $\uparrow$} & {FSNR} & {CD $\downarrow$} &  {SRMR $\uparrow$} & {DNSMOS $\uparrow$} & {WER $\downarrow$} \\ 
         \midrule         
	  Input & 1.12 &	0.56 &	1.44 &	5.62 &	3.11 &	2.50 &	63.47 / 41.32 
\\
         \midrule
         DFN3 \cite{schroter23b_interspeech} 
         & \bfseries 1.23 &	0.61 &	\bfseries 1.99 &	4.46 &	5.90 &	2.65 &	67.98 / 41.01 \\
          \multicolumn{1}{r}{+ FiLM} 
          & \bfseries 1.23 &	\bfseries 0.62 & 1.94 &	\bfseries 4.34 &	\bfseries 6.14 &	\bfseries 2.66 &	\bfseries 67.81 / 39.41 \\ 
         \midrule
         Two-step \cite{rosenbaum_2025}
          & 1.29 &	0.61 &	1.63 &	4.50 &	7.24 &	2.64 &	68.90 / 42.08 \\ 
         \multicolumn{1}{r}{+ FiLM}
         & \bfseries 1.31 &	\bfseries 0.64 &	\bfseries 1.89 &	\bfseries 4.28 &	\bfseries 8.20 &	\bfseries 2.72 &	\bfseries 64.61 / 35.15 
         \\
        \bottomrule
    \end{tabular}
    
    \label{tab:joint_results}
    \vspace{1.25mm}
\end{table}

\subsection{Enhancement results}
The results for speech enhancement performance are shown in \Cref{tab:joint_results}.
As can be seen from the metrics on the input mixture signals, the data pose challenging conditions.
For reverberant data, conditioning DFN3 and the two-step approach leads to small but consistent improvements in all metrics. Therefore, additional information on the \glspl{RIR} encoded in the embeddings leads to more robust estimates of the filter coefficients that improve dereverberation performance compared to the non-conditioned models.
Note that the \gls{WER} results for the weak ASR model on the enhanced results even degrade compared to the input. The chosen deep filtering approach seems to introduce artifacts that the weak ASR system cannot handle, although dereverberation is applied, as seen in the other metrics. However, the use of additional conditioning leads to improvements compared to the non-conditioned systems, and for the strong \gls{ASR} system it also improves upon the input signal \gls{WER}.

For noisy-reverberant data, the results are similar, with consistent improvements for FiLM conditioning in all metrics. Interestingly, the gain through FiLM conditioning is even larger for the strong ASR model than for the weak one.

\section{Conclusions}
\label{sec:conclusion}
In this work, we developed an approach to train noise-robust \gls{RIR} embeddings. We adopted a three-stage approach, where the model is first trained in a contrastive framework on reverberant data, then on noisy and reverberant data, before finally a teacher-student approach is used, where the student is tasked with deriving the teacher embeddings from the noisy version of the reverberant input. Evaluations on the estimation of \tsix and \gls{DRR} showed that these parameters are linearly accessible from the embeddings and are maintained by the student model for both reverberant and noisy-reverberant input. The conditioning of deep filtering-based speech enhancement models on the RIR embeddings resulted in consistent gains in all considered metrics, including the word error rate of a weak and a strong ASR model. As future work, we are planning to investigate other discriminative approaches and multi-stage models. Further, we aim to extend the \gls{RIR} representation to multi-channel input and frame-wise resolution for use in speech enhancement in dynamic scenarios.

\clearpage

\bibliographystyle{IEEEbib}
\bibliography{strings,refs}

\end{document}

%% file: macros.tex
\newcolumntype{H}{>{\setbox0=\hbox\bgroup}c<{\egroup}@{}}   %

\glsdisablehyper    %
\newacronym{RIR}{RIR}{room impulse response}
\newacronym{ASR}{ASR}{automatic speech recognition}
\newacronym{DRR}{DRR}{direct-to-reverberant energy ratio}
\newacronym{SNR}{SNR}{signal-to-noise ratio}
\newacronym{SSL}{SSL}{self-supervised learning}
\newacronym{WER}{WER}{word error rate}
\newacronym{MAE}{MAE}{mean absolute error}

\usepackage{pgfplots}
\pgfplotsset{compat=1.18}
\usepgfplotslibrary{groupplots}

\usetikzlibrary{shadings, calc, matrix}
\usepackage{twemojis}

\pgfplotsset{
    colormap={mymap}{
        rgb255=(31,119,180);  %
        rgb255=(255,127,14);  %
        rgb255=(44,160,44);   %
        rgb255=(214,39,40);   %
        rgb255=(148,103,189); %
        rgb255=(23, 190, 207) %
        rgb255=(127,127,127); %
        rgb255=(188,189,34);   %
        rgb255=(227, 119, 194); %
    }
}

\definecolor{tabblue}{RGB}{31,119,180}
\definecolor{taborange}{RGB}{255,127,14}
\definecolor{tabgreen}{RGB}{44,160,44}
\definecolor{tabred}{RGB}{214,39,40}
\definecolor{tabpurple}{RGB}{148,103,189}
\definecolor{tabcyan}{RGB}{23,190,207}
\definecolor{tabgrey}{RGB}{127,127,127}
\definecolor{tabolive}{RGB}{188,189,34}
\definecolor{tabpink}{RGB}{227, 119, 194}

%% file: images/teacher_student.tex
\begin{tikzpicture}[
    line width=1.15pt,
    studentcolor/.style={
        draw=orange!75!black,
        fill=orange!13,
        text=orange!40!black
    },
    teachercolor/.style={
        draw=cyan!60!blue,
        fill=cyan!10,
        text=cyan!45!blue
    },
    featurecolor/.style={
        draw=gray!55!black,
        fill=gray!7,
        text=gray!45!black
    },
    losscolor/.style={
        draw=gray!55!black,
        fill=gray!7,
        text=gray!45!black
    },
    block/.style={
        draw,
        rounded corners=9pt,
        line width=1.25pt,
        minimum width=2cm,
        minimum height=1.cm,
        align=center,
        drop shadow={
            shadow xshift=0.7pt,
            shadow yshift=-0.7pt,
            opacity=0.16
        }
    },
    sum/.style={
        draw=gray!70!black,
        fill=white,
        line width=1.15pt,
        circle,
        minimum size=9mm,
        inner sep=0pt
    },
    signal/.style={
        ->,
        line width=1.1pt,
        draw=gray!80!black
    }
]

\node (teacher) [block, teachercolor, align=center] {%
    \textbf{Teacher} \\[ 1.ex] \large \texttwemoji{snowflake}
};

\node (student) [
    block,
    studentcolor,
    below=0.75cm of teacher,
    align=center
] {%
    \textbf{Student} \\[ 1.ex] \large{\texttwemoji{fire}}
};

\node (mse) [
    block,
    losscolor,
    minimum width=2.1cm,
    right=1.5cm of $(teacher)!0.5!(student)$
] {%
    $\mathcal{L}_\text{MSE}$
};

\node (features) [
    block,
    featurecolor,
    left=1.75cm of teacher,
    minimum width = 1cm,
] {%
    $h_1$\\[3pt]
    $h_2$
};

\node (u1) [
    left=0.5cm of features.north west,
    yshift=-0.35cm
] {$u_1$};

\node (u2) [
    left=0.5cm of features.south west,
    yshift=0.35cm
] {$u_2$};

\draw[signal] (u1.east) -- (features.west |- u1.east);
\draw[signal] (u2.east) -- (features.west |- u2.east);

\coordinate (branch) at ($(features.east)!0.40!(teacher.west)$);

\draw[signal]
    (features.east)
    -- node[midway, above=3pt]
       {$s_1,\,s_2,\,s_3$}
    (teacher.west);

\node (sum) [sum] at (branch |- student.center) {$+$};

\node (noise) [left=0.5cm of sum] {$n$};
\draw[signal] (noise.east) -- (sum.west);

\draw[signal] (branch) -- (sum.north);

\draw[signal] (sum.east) -- (student.west);

\coordinate (mseTopIn)    at ($(mse.north) + (0, 0.cm)$);
\coordinate (mseBottomIn) at ($(mse.south) + (0, 0.cm)$);

\coordinate (teacherBend) at ($(mseTopIn) + (0cm,0)$);
\coordinate (studentBend) at ($(mseBottomIn) + (0cm,0)$);

\draw[signal]
    (teacher.east)
    -- node[midway, above right=3pt and -18pt, text=cyan!50!blue]
       {$\mathbf{e}_1,\,\mathbf{e}_2,\,\mathbf{e}_3$}
       (teacherBend |- teacher.east)
    -- (mseTopIn);

\draw[signal]
    (student.east)
    -- node[midway, below right=3pt and -18pt , text=orange!50!black]
       {$\mathbf{\hat{e}}_1,\,\mathbf{\hat{e}}_2,\,\mathbf{\hat{e}}_3$}
       (studentBend |- student.east)
    -- (mseBottomIn);

\fill[blue!55!black] (branch) circle (1.6pt);

\end{tikzpicture}